# A BCI based Smart Home System Combined with Event-related Potentials and Speech Imagery Task

Hyeong-Jin Kim[1], Min-Ho Lee[2], Minji Lee[1]
[1]Department of Brain and Cognitive Engineering, Korea University, Seoul, Republic of Korea
[2]Department of Computer Science, Nazarbayev University, Nur-Sultan, Kazakhstan
kme0115@korea.ac.kr, minho.lee@nu.edu.kz, minjilee@korea.ac.kr

*Abstract*—Recently, smart home systems based on brain-computer interface (BCI) has attracted a wide range of interests in both industry and academia. However, the current BCI system has several shortcomings as it produces a comparatively lower accuracy for real-time implementations as well as the intuitive paradigm for the users cannot be well established here. Therefore, in this study, we proposed a highly intuitive BCI paradigm that combines event-related potential (ERP) with the speech-imagery task for the individual target objects. The decoding accuracy of the proposed paradigm was 88.1% ($\pm 5.90$) which is a much significant higher performance than a conventional ERP system. Furthermore, the amplitude of N700 components was significantly enhanced over frontal regions which are priory evoked by the speech-imagery task. Our results could be utilized to develop a smart home system so that it could be more user-friendly and convenient by means of delivering user's intentions both, intuitively and accurately.

## I. INTRODUCTION

The brain-computer interface (BCI) allows people to perform interactions with other people and devices (from simple switches to complex wheelchairs or smart homes) without activating muscles [1-4]. In particular, a system based on electroencephalography (EEG) is a non-invasive interface that utilizes brain signals from the scalp to determine the user's intentions [5-7]. It has the advantage of having a low price, flexibility, and high in time resolution [8-10]. Therefore, BCI is as considered one of the next-generation interfaces and even people with physical disabilities can control objects in smart homes without moving their muscles [11].

Event-related potential (ERP) paradigm has been used extensively as one of the major components in the speller system [12]. This paradigm has several advantages such as high performance in a high degree of freedom and easy to train even for novice subjects [13, 14]. Many studies have reported smart home control using ERP-based BCI [15, 16]. Edlinger et al. [15] first applied the conventional ERP speller paradigm to smart home control. The average performance was 94.0% in the three subjects. However, each subject performed different tasks and the number of subjects was too small to analyze in statistics. Therefore, they suggested only the possibility of smart home control based on BCI but is not possible to use for real-life applications. Zhang et al. [16] proposed another environmental control system using ERP. This study focused on patients with severe spinal cord injuries. They showed every patient could use the ERP-based environmental control system satisfactorily. However, this method has only 25 selectable objects and needs an additional device. Also, this takes much time because it was based on a single character speller system [17]. In this regard, it needs a method to take less time while retaining high accuracy.

To compensate for the limitations of ERP, a new paradigm has been applied for controlling the smart home. The steady-state visual evoked potential (SSVEP) is one of the major EEG paradigms which uses the phenomenon of synchronization to the frequency of the brain [18-20]. For example, evoked brain responses at 12 Hz are generated in the occipital lobe, which is the primary visual cortex, when presenting a visual stimulus of 12 Hz [20]. Gao et al. [21] suggested a smart home control system using SSVEP. It has the advantage of short training time and a high signal-to-noise ratio. The average performance was 93.2% on four selectable objects during 4 seconds. However, this paradigm has a fatal limitation which could cause eyestrain, even epileptic seizures [18-20]. In addition, so far, smart home control systems based on ERP or SSVEP are not intuitive and have disadvantages that users cannot use them comfortably.

In order to solve the above problems and to present a more intuitive paradigm, we additionally used a relatively recent method called imagined speech. This paradigm indicates a mental process of imagining the utterance of a word without emitting sounds or articulating any movement [22]. In addition, this is associated with Broca's area where the left inferior frontal lobe, Wernicke's area where the left superior temporal lobe, and the motor cortex [23]. Specifically, in this paradigm, the speech-related potentials peak is at 350 ms after the onset of the speech, enabling smart home control by utilizing this feature [24, 25]. Therefore, this paradigm is likely to be used as the intuitive signals for decoding the thoughts themselves as the most direct type of communication interface [26]. However, the imagined speech is low in performance for direct use in real life.

In this study, we proposed a smart home control system using a fresh paradigm that combines ERP and the imagined speech. In particular, we divided the imagined speech into two

This work was partly supported by Institute for Information & Communications Technology Planning & Evaluation (IITP) grant funded by the Korea government (MSIT) (No. 2015-0-00185, Development of Intelligent Pattern Recognition Softwares for Ambulatory Brain-Computer Interface, No. 2017-0-00451, the Development of BCI-based Brain and Cognitive Computing Technology for Recognizing User's Intentions using Deep Learning) parts: imagining meaningful words or not. We used two major components of the ERP paradigm. P300 is a strong positive brain response that is extracted approximately 300 ms after the onset of the stimulus [27]. Most studies using ERP exploit the P300 component for analysis. This is because it shows the most remarkable difference between the target and the non-target [28]. N700, like the P300, one of a strong negative brain response that is extracted approximately 700 ms after the onset of the stimulus [29]. We used this component because it also represented a remarkable difference between the target and the non-target in imagined speech. We hypothesized that the performance in the combined paradigm using imagining meaningful words was higher than in the conventional paradigm using ERP. In addition, we have implemented a virtual smart home that can control 36 objects which could be applied in real-life. These results would be able to provide insight into smart home control based on BCI.

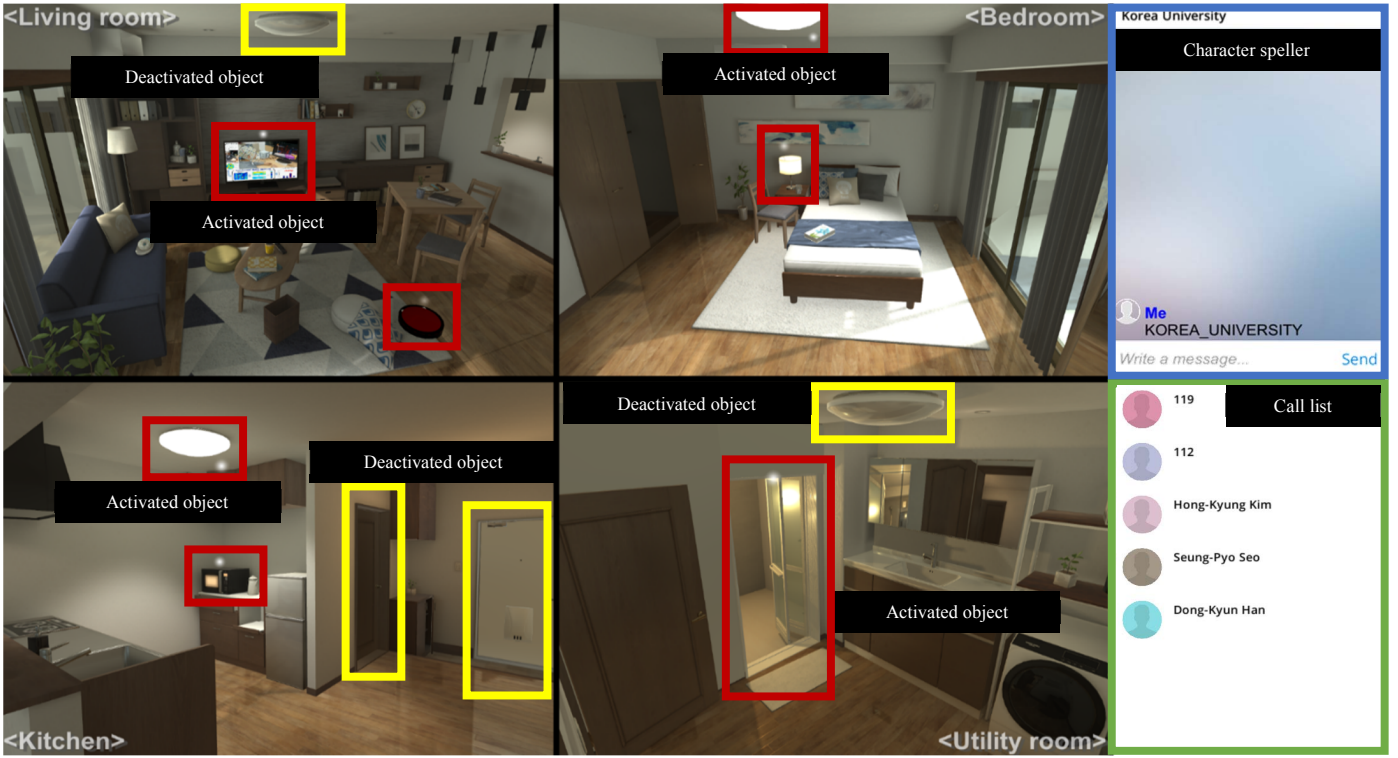

Figure 1. The virtual smart home system which can interact with the user. The red boxes represent the object activated by the user and the yellow box shows deactivated objects. The blue box denotes space for the result of the character speller and the green box displays an able call list.

## II. MATERIALS AND METHODS

### A. Data Acquisition

Eight healthy novice subjects who were between 22 and 27 years (two females, average age 23.9) participated in our study. The subjects were seated in a comfortable chair at approximately 60 cm from the screen using a 5.7-inch smart-phone (60 Hz refresh rate, 1920×1080 resolution). During the experiment, they could stop and rest if desired. Also, they were asked to avoid unnecessary movements.

EEG activity was acquired from 32 channels using ActiCap EEG amplifier (Brain Product; Munich, Germany) in 100 Hz sampling rate. The tip of the nose and the mastoid region of ear was selected as the location of the reference and the ground electrode. This study was reviewed and approved by the Korea University Institutional Review Board (1040548-KUIRB-16-159-A-2), and written informed consent was obtained from all subjects before the experiments.

### B. Experimental Procedures

We developed the virtual smart home in Unity (Unity Technologies, CA, USA) (Fig. 1). There are four rooms and the selectable target objects are 36 in this virtual smart home. Each object has a one-to-one correspondence with the images displayed in the smart home speller. When one of them is selected, it performs the specified action (e.g. open and close the door or turn on and off the light). Also, the users can type the alphabet, number (1-9), and underline (_) if they select the icon representing the green dashed box.

Fig. 2 (a) indicates the main monitor shown to the subject. If the target image is not highlighted, the subject waits for the target to be highlighted. When the target image is highlighted, the subject performs the appropriate action requested for each condition. The interface layout was designed with 36 target visual stimuli. The stimuli were equally spaced on the screen in this configuration.

Fig. 2 (b) shows the overall procedure for each trial. Each trial is composed of 10 sequences and each sequence consists of 12 flashing stimuli. Every flash contains 6 objects in random order. Every object blinked twice with the white color in a sequential manner. Each stimulus was given for 50 ms and the inter-stimulus interval is given for 135 ms. The presentation order was randomized, and non-target stimuli were flashed consecutively in order to avoid the repeated flashing of the same target stimulus between the sequences. The paradigm code was developed with the Psychophysics Toolbox and OpenBMI [30] in Matlab (MathWorks, MA, USA).

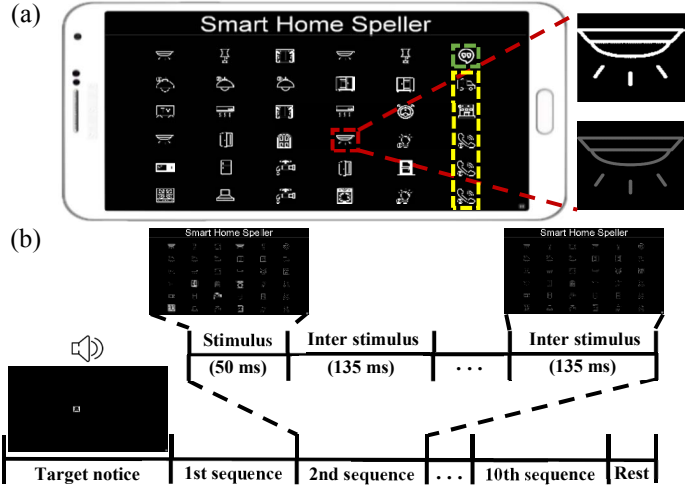

Figure 2. The screen to display for the subject and overall procedure for each trial. (a) The large pictures which are expanded with a red dashed line represent the image whether it is highlighted and not. The green dashed box changes the smart home speller screen to the character speller screen. And the yellow dashed box has correspondence with the call list. (b) The overall procedure for each trial. The subject needs 10 sequences to select a target object.

We designed three conditions as follows: (i) ERP only, (ii) combined stimuli: ERP + imagined speech for saying meaningless word, and (iii) combined stimuli: ERP + imagined speech for saying meaningful word. In the first condition, each subject was requested to react to the visual stimuli. In the second condition, the subject is demanded to imagine saying a meaningless word while inducing a response to the stimulus. In our case, we chose the word 'Ah'. And in the last third condition, the subject is requested to imagine speaking the word of the target object name while inducing the response to the stimulus. We experimented twice with the training phase and the testing phase. In each phase, the subject repeated 20 times the task a selecting one of the 36 objects.

### C. EEG Data Analysis

The EEG data were segmented from -200 to 800 ms regarding the stimulus onset. Also, they were baseline-corrected by subtracting the mean amplitudes in the -200 to 0 ms pre-stimulus interval from the epoch. Mean amplitude values were calculated for all channels in each selected time window. Therefore, these subject-dependent spatio-temporal feature vectors were formed with 320 dimensions (i.e., 32 channels × 10 features). We used the regularized linear discriminant analysis (RLDA) [31] to classify the brain signals. It was trained using these feature vectors from the target and non-target trials [14, 32]. The data from the training phase was used for training the classifier and the data from the testing phase was used as a validation set.

The L2 norm value between the target and non-target amplitude was calculated to figure out the remarkable difference at each component. We chose this because it is the most widely used method to measure the magnitude of the vector in the Euclidean vector space [33].

$$\|x\|_2 := \sqrt{\sum_{i=1}^{n}(t_i - nt_i)^2} \quad (1)$$

where $t_i$, $nt_i$ means $i$ th target and non-target sample's amplitude, and $n$ denotes the number of the samples. In the case of P300, the analysis was performed using the mean waveforms from 200 ms to 400 ms after the stimulus was given. Also, N700 was analyzed using the mean waveforms from 600 ms to 800 ms after the onset of the stimulus.

### D. Statistical Analysis

We performed Kruskal-Wallis tests for the classification performance in each sequence in three conditions. In addition, Kruskal-Wallis tests were applied for the L2 norm value of P300 and N700 between the target and non-target in three conditions. Also, Wilcoxon rank-sum tests were performed for a post-hoc analysis. In the statistical analysis, the significance level was set at $\alpha = 0.05$ with Bonferroni correction.

## III. RESULTS

### A. Classification Performance

We explored to select 36 target objects in each condition at the virtual smart home. There are significant differences in classification performance among three conditions except for 1, 2, and 8 sequences (Table I). Fig. 3 denotes the mean accuracy according to the sequence of each condition. The highest accuracy in condition 1 is 59.4 ± 28.8% at the sixth sequence, 77.5 ± 19.6% in condition 2 for the same sequence, and 88.1 ± 5.9% at the seventh sequence in condition 3, respectively. The classification accuracy in condition 2 was higher than in condition 1 (third sequence: $p = 0.019$, fourth sequences: $p = 0.034$). The classification performance in condition 3 was also higher than in condition 1 (fourth sequence: $p = 0.003$, fifth sequence: $p = 0.002$, sixth sequence: $p = 0.010$, seventh sequence: $p = 0.002$, ninth sequence: $p = 0.009$, tenth sequence: $p = 0.001$). However, there was no significant difference in classification accuracy between condition 2 and condition 3.

TABLE I. STATISTICAL RESULTS FOR MEAN ACCURACY ACCORDING TO THE SEQUENCE

| Sequence | Classification accuracy | |
|---|---|---|
| | Chi-square | p-value |
| 1 | 1.090 | 0.580 |
| 2 | 0.378 | 0.828 |
| 3 | 6.348 | **0.042** |
| 4 | 8.432 | **0.015** |
| 5 | 9.383 | **0.009** |
| 6 | 6.550 | **0.038** |
| 7 | 7.592 | **0.022** |
| 8 | 5.629 | 0.060 |
| 9 | 6.501 | **0.039** |
| 10 | 8.853 | **0.012** |

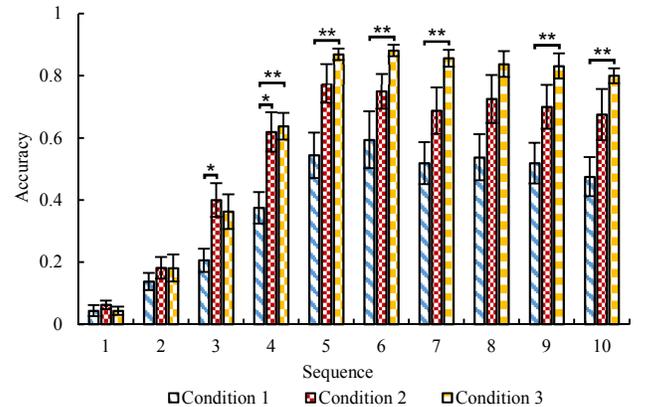

Figure 3. Mean accuracy according to the sequence of each condition. * means significantly different with no Bonferroni correction and ** demonstrates significantly different with Bonferroni correction.

TABLE II. STATISTICAL RESULTS FOR P300 PEAK IN EACH CHANNEL

| P300 ERP peak | | | | | | | | | | | |
|---|---|---|---|---|---|---|---|---|---|---|---|
| *Frontal* | | | *Central* | | | *Parietal* | | | *Occipital* | | |
| *Channel* | *Chi-square* | *p-value* | *Channel* | *Chi-square* | *p-value* | *Channel* | *Chi-square* | *p-value* | *Channel* | *Chi-square* | *p-value* |
| Fp1 | 24.160 | **<0.001** | C3 | 2.844 | 0.241 | FC5 | 4.545 | 0.103 | P3 | 0.831 | 0.660 |
| Fp2 | 21.783 | **<0.001** | C1 | 1.103 | 0.576 | FC6 | 6.360 | **0.042** | P1 | 2.032 | 0.362 |
| F3 | 6.849 | **0.033** | Cz | 1.548 | 0.461 | T7 | 6.709 | **0.035** | Pz | 1.171 | 0.557 |
| Fz | 4.399 | 0.111 | C2 | 2.671 | 0.263 | T8 | 11.243 | **0.004** | P2 | 1.556 | 0.459 |
| F4 | 5.859 | 0.053 | C4 | 3.484 | 0.175 | CP5 | 0.667 | 0.717 | P4 | 3.047 | 0.218 |
| FC1 | 2.801 | 0.246 | CP1 | 0.715 | 0.699 | CP6 | 4.558 | 0.102 | O1 | 1.316 | 0.518 |
| FCz | 1.931 | 0.381 | CPz | 0.894 | 0.640 | P7 | 5.189 | 0.075 | Oz | 4.525 | 0.104 |
| FC2 | 2.219 | 0.330 | CP2 | 0.742 | 0.690 | P8 | 1.397 | 0.497 | O2 | 5.202 | 0.074 |

TABLE III. STATISTICAL RESULTS FOR N700 PEAK IN EACH CHANNEL

| N700 ERP peak | | | | | | | | | | | |
|---|---|---|---|---|---|---|---|---|---|---|---|
| *Frontal* | | | *Central* | | | *Parietal* | | | *Occipital* | | |
| *Channel* | *Chi-square* | *p-value* | *Channel* | *Chi-square* | *p-value* | *Channel* | *Chi-square* | *p-value* | *Channel* | *Chi-square* | *p-value* |
| Fp1 | 26.589 | **<0.001** | C3 | 37.198 | **<0.001** | FC5 | 35.963 | **<0.001** | P3 | 11.958 | **<0.001** |
| Fp2 | 22.641 | **<0.001** | C1 | 34.573 | **<0.001** | FC6 | 35.034 | **<0.001** | P1 | 10.960 | **<0.001** |
| F3 | 34.993 | **<0.001** | Cz | 32.010 | **<0.001** | T7 | 25.414 | **<0.001** | Pz | 7.761 | **<0.001** |
| Fz | 37.562 | **<0.001** | C2 | 35.144 | **<0.001** | T8 | 33.412 | **<0.001** | P2 | 13.844 | **<0.001** |
| F4 | 32.538 | **<0.001** | C4 | 38.738 | **<0.001** | CP5 | 33.769 | **<0.001** | P4 | 14.507 | **<0.001** |
| FC1 | 37.940 | **<0.001** | CP1 | 27.185 | **<0.001** | CP6 | 35.801 | **<0.001** | O1 | 2.514 | 0.285 |
| FCz | 40.999 | **<0.001** | CPz | 23.426 | **<0.001** | P7 | 12.959 | **<0.001** | Oz | 0.264 | 0.876 |
| FC2 | 40.654 | **<0.001** | CP2 | 21.830 | **<0.001** | P8 | 18.000 | **<0.001** | O2 | 3.037 | 0.219 |

*B. ERP Peak*

We investigated the differences in the P300 and N700 components in all conditions. For the P300 component, there are only 6 channels (Fp1, Fp2, F3, FC6, T7, and T8) which have a significant difference in the L2 norm value for the amplitude of target and non-target between each condition. Although there are significant differences in amplitude between each condition in the six channels, none of the channels show significant differences in L2 norm value for the amplitude of target and non-target between all conditions when comparing conditions individually (Table II).

Table III represents the statistical results for N700 in each channel. In contrast to the P300, N700 shows every channel has a significant difference in L2 norm value for the amplitude of target and non-target between each condition except to the O1, Oz, and O2. In addition, it is mainly the 10 channels in the frontal area that shows the difference between each of the three conditions. Fig. 4 shows the L2 norm value for the amplitude of target and non-target between each condition in frontal regions. In the 10 channels (F3, Fz, F4, FC5, FC1, FCz, FC2, FC6, T7, and C3), the N700 of condition 3 was the highest, the N700 of condition 1 was the lowest, and the condition 2 had the value between them statistically.

IV. DISCUSSION AND CONCLUSION

In this study, we proposed a fresh smart home control system using ERP and imagined speech. The combined ERP and imagined speech for object name statistically represented higher performance compared to the conventional ERP paradigm. We also found statistical differences between the three conditions in the N700 in the frontal area. N700 was the highest in condition 3 and lowest in condition 1. It is considered that these characteristics may have affected the classification performance for controlling a virtual smart home.

We observed that N700 peak in condition 3 was higher than in condition 1. However, no significant differences in P300 peak between condition 1 and condition 3 were found. In the previous ERP paradigm, P300 is the main feature compared to N700 [27], whereas N700 is more prominent than P300 when speech is concerned [34]. These features were especially prominent in the frontal region when the user spoke [34]. In addition, the characteristics of N400 appear in the frontal region in speech task [35]. Importantly, similar brain changes occur when we speak or imagine speech [36]. In this regard, it is considered that the same characteristics such as the N400 or N700 appear when performing an imagined speech like actual speech. Therefore, in condition 3, there was no significant difference in P300; because the features were offset by changes in P300 and N400. On the other hand, the differences in N700 between condition 1 and condition 3 were starkly observed; because the features in N700 were probably increased in condition 3 due to the imagined speech.

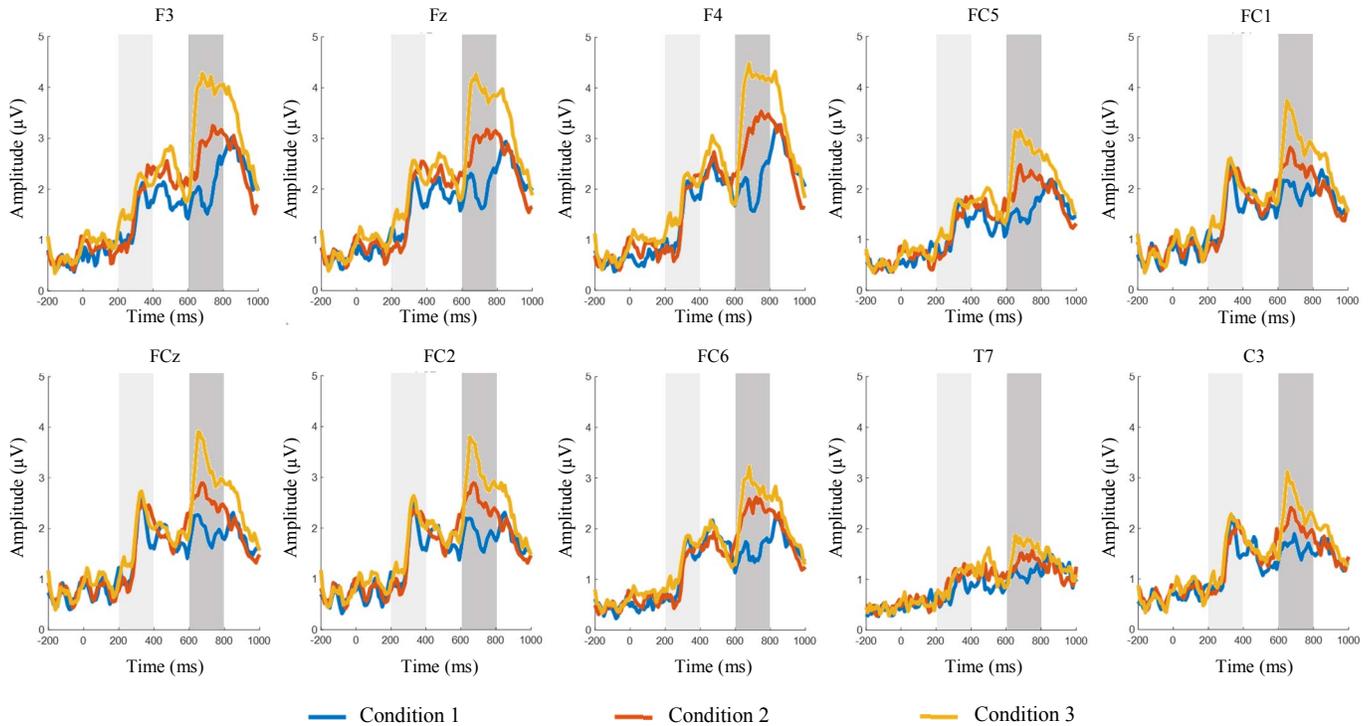

Figure 4. L2 norm value for the amplitude of target and non-target between each condition according to the channels. The bright-gray area and dark-gray area section were extracted to analyze the P300 and N700 peak each. The selected channels show a significant difference in amplitude between all conditions.

This study has some limitations. Firstly, we did not observe significant differences in the classification performance between condition 2 and condition 3. However, the N700 peak in condition 3 was clearly higher than in condition 2. Therefore, when comparing performance with more subjects, it is expected that condition 3 could be higher, which had been imagined as saying something meaningful. Secondly, we just trained the classifier with a simple machine learning model, RLDA. However, the classification accuracy of the classifier trained with a deep learning model that can reflect the features of the data to be much higher than our current result [37, 38]. In addition, if we applied K-fold cross validation, which is widely used in research using biological signals that are difficult to obtain a lot of data we would have had higher performance. Thirdly, classification performance in condition 1 was lower than conventional studies using ERP. However, Obeidat et al. [39] showed the decrease of performance in smartphone compared to computer monitor because of the size of the stimulus and the interval between each target. Therefore, our low performance is not a problem and our classification performance using only ERP is similar to Obeidat et al. [39] in a smartphone environment. However, in order to utilize the practical smartphone ERP speller, performance should be improved. Lastly, the proposed method works according to visual and auditory stimuli, so this method cannot be used if the user has problems with the sensory organs. In the future, smart home control systems will need to be developed for these patients as well.

The paper concludes stating that, the proposed method combining ERP with the imagined speech showed statistical superiority over the conventional ERP paradigm in the classification accuracy. This is because the P300 component has similar representations in both methods, but the N700 peak has shown significantly higher improvement in the proposed method. The proposed method might replace the widely used conventional ERP paradigm for controlling a smart home.

ACKNOWLEDGMENT

We thank H.-K. Kim for helping to implement the smart home control system.